\documentstyle{mn}

\begin{document}

\title[Cosmic strings with self-interacting hot dark matter]
{Cosmic strings with interacting hot dark matter}

\author[P.\ P.\ Avelino and J.\ P.\  M.\ de~Carvalho]
  {P.\ P.\ Avelino$^{1}$\thanks{E-mail: pedro@astro.up.pt} and J.\ P.\  M.\
de~Carvalho$^{1,2}$\thanks{E-mail: 
mauricio@astro.up.pt} \\
  $^{1}$Centro de Astrof{\' \i}sica da Univ.\ do Porto, Rua do Campo Alegre 823, 4150 
Porto -- Portugal \\
  $^{2}$Dep.\ de Matem\'atica Aplicada da Faculdade de Ci\^encias da Univ.\ do Porto,
Rua das Taipas 135, 4050  Porto -- Portugal}

\date{}

\maketitle

\begin{abstract}

We compute the linear power spectrum of cosmic string sedeed fluctuations in the context of neutrinos with 
a strong self-interaction and show that it is very similar to that obtained in the context of `normal' 
neutrinos. We compare our results with observational data and show that for any value of the 
cosmological parameters $h$ and $\Omega_0$ the interacting hot dark matter power spectrum 
requires a large scale dependent biasing parameter.

\end{abstract}


\section{Introduction}

Observations on all scales of cosmological interest show that most of the
matter in the universe is in some form of unseen matter. On the other
hand, nucleosynthesis constraints seem to imply that most of this dark matter is non
baryonic. Gravitinos, axions and neutrinos for example, are only a few among the many 
candidates to constitute the dark matter.

In the standard scenario it is usually assumed that during the important epochs
for structure formation, the dark matter particles candidates
interact only gravitationally. However one could assume that this is not the
case. In particular it is possible that light
neutrino-like particles could have some other kind of self-interaction during
the low energy epochs when structure formation takes place (see, for instance,
Raffelt and Silk 1987; Carlson, Machacek and Hall 1992; Gradwohl and Frieman
1992; Machacek 1994; Laix, Scherrer and Schaefer 1995).

In recent work Atrio-Barandela and Davidson (1997), ha\-ve considered a light
($\sim 30\,$eV) self-interacting neu\-tri\-no-like particle and discussed the
possibility that the dark matter in the universe could be constituted by this
kind of particles. They determined the linear power spectrum of density
fluctuations generated by the present time in the context of primordial gaussian
fluctuations and concluded that galaxy sized density perturbations could
survive.

In what concerns the estimates for the bounds on the neutrino-neutrino cross
section very little is known. Within the framework of the Standard Model of
Particle Physics, no experimental bounds have been set, and the best known
lower limit on the mean free path of the neutrinos is based on observations of
the supernova $SN\,1987\,A$. In fact, the neutrinos produced by $SN\,1987\,A$ in
the Large Magellanic Cloud, reached the earth after crossing the relic
neutrinos of the cosmic background. Since the number of detected neutrinos
agrees with the theoretical expected number from supernova models, the present mean free
path of the neutrinos $\lambda_0$, is not shorter than the distance to $SN\,1987\,A$,
which is of the order of $2\times10^{23}\,$cm; for a complete analysis, see Kolb and
Turner, 1987; (see also Nussinov and Roncadelli, 1983). In this article we will assume
that the co-moving value of $\lambda$ is much smaller than any scale of cosmological
interest during the time where most of the
structures we observe in the universe today are generated. In this case one can
treat the neutrino component with the hydrodynamic or fluid approximation.

Here we extend the work of Barandela and Davidson (1997) to the
case of non-gaussian density perturbations induced by cosmic strings. Cosmic
strings, along with other topological defects, may be produced as a result of symmetry breaking
phase transitions in the early universe (Kibble 1976). Although some topological
defects are ruled out by observations, others like cosmic strings could be the seeds 
which generated the large scale structures we observe in the universe today. They
also have other interesting cosmological consequences (see for example Vilenkin 1985;
Colombi 1993; Vilenkin and Shellard 1994; Avelino 1996). Among these there is a
non-gaussian signature in the microwave background (on small scales), the production of
double images and a primordial background of gravitational waves or other radiation.

The outline of this paper is as follows. In section $2$ we start by describing the formalism we employ in order to study
the growth of density fluctuations in the context of neutrinos with a strong self-interaction. We introduce compensation
in section $3$ and in section $4$ we describe the semi-analitic model we employ in order to compute the power spectrum of
cosmic string seeded density fluctuations. In section $5$ we describe and discuss the results. We conclude in section $6$.

In this article employ fundamental units with $\hbar = c = k_B = 1$. We consider a 
self-interacting neutrino with mass $m_\nu = 93 \Omega_0 h^2$ such that $\Omega_0=\Omega_\nu$.

\section{Subsequent perturbations} \label{sec.6.1}

In a flat FRW universe with no cosmological constant and containing both CDM and
radiation fluids, the scale factor $a$ may be written in terms of the conformal
time $\eta$ as
\begin{equation}  
a(\eta) = a_{eq} \Big[ 2(\sqrt{2} - 1) {\eta \over \eta_{eq}} + (3 - 2
\sqrt{2}){\eta^2 \over \eta_{eq}^2} \Big], \label{one}
\end{equation}
where $a_{eq}$ and $\eta_{eq}$ are respectively the expansion
factor and conformal time  at matter-radiation equality. If the dark matter is
hot but the transition from the relativistic to the non-relativistic regime
occurs deep in the radiation era eqn. (\ref{one}) is still a good
approximation. 

The evolution of radiation and non-relativistic interacting hot dark matter (IHDM) density
fluctuations in the synchronous gauge is given by
\begin{eqnarray} \ddot \delta_h + {\dot a \over a} \dot \delta_h - {3 \over
2}\Big({\dot a \over a}\Big)^2 (\Omega_h \delta_h +  2 \Omega_r \delta_r) -
c_s^2 \nabla^2 \delta_h \nonumber \\ 
= 4 \pi (\Theta_{00} +
\Theta_{ii}),
\label{two}
\end{eqnarray}
\begin{equation} \ddot \delta_r - {1 \over 3} \nabla^2 \delta_r  - {4 \over 3}
\ddot \delta_h = 0,
\label{three}
\end{equation}
where $\Theta_{\alpha\beta}$ is the energy-momentum tensor of the external
source, $c_s$ is the adiabatic sound speed of the interacting hot dark
matter and $\Omega_h$ and $\Omega_r$ express the densities in radiation and IHDM as fractions of
the critical density.
 
We are implicitly assuming that the strength of the neutrino-neutrino coupling
is large enough for it to be a good approximation to treat the neutrino
component as a perfect fluid. We also assume the radiation behaves as a fluid;
this is certainly valid before recombination for the length scales  of
interest, well above the radiation damping scale. Although this is not true
after recombination, for $\Omega_0 h^2$ not too small the radiation no longer
dominates the dynamics after last scattering. Consequently, we expect that even
then equation (\ref{three}) remains a valid approximation.
 
The solution to the system of equations (\ref{two},\ref{three}) with initial conditions 
$\delta_h =\delta_r={\dot \delta_h}={\dot \delta_r}=0$ can be written in terms of Green functions
as
\begin{equation}
\delta^S_{h,r}({\bf x},\eta) = 4 \pi \int_{\eta_i}^\eta d\eta' \,
\int d^3x' {\cal G}^{h,r}(X;\eta,\eta') \Theta_{+}({\bf x'},\eta'),
\end{equation}
\begin{equation}
{\cal G}^{h,r}(X;\eta,\eta') = {1 \over 2 \pi^2} \int_0^\infty \, \widetilde {\cal
G}^{h,r}(k;\eta,\eta') {\sin k X \over k X} k^2 dk. 
\end{equation} 
Here, $\Theta_{+} \equiv \Theta_{00}+\Theta_{ii}$ and $X=|{\bf x} -{\bf x'}|$. The upper index `S'
indicates that these are the `subsequent' fluctuations, according to the
notation of Veeraraghavan \& Stebbins (1990), to be distinguished from the
`initial' fluctuations. In Fourier space, the Green function $\widetilde {\cal
G}^{h,r}$  obey the same equations as
$\delta_{h,r}$:
\begin{equation} \ddot{\widetilde {\cal G}^h} + {\dot a \over a}
\dot{\widetilde {\cal G}^h} - {3 \over 2}\Big({\dot a \over a}\Big)^2 \, (
\Omega_h \widetilde {\cal G}^h + 2 \Omega_r  \widetilde {\cal G}^{r}) + c_s^2 k^2 {\widetilde
{\cal G}^h}= 0, \label{gcdmeqn} 
\end{equation} 
\begin{equation}  
\ddot{\widetilde {\cal G}^{r}} - {4 \over 3} 
\ddot{\widetilde {\cal G}^h} + {1 \over 3} k^2 
\widetilde {\cal G}^{r} = 0. \label{gradeqn}
\end{equation} 
The initial conditions at $\eta_i$ are  $\dot{\widetilde {\cal G}^h} = 1$, 
$\dot{\widetilde {\cal G}^r} = 4/3$ and $\widetilde {\cal G}^{h} = \widetilde
{\cal G}^{r} = 0$. Hence, given the evolution of the source stress-energy, it
is possible to numerically compute the resulting density fluctuations.

We are mostly interested in the interacting IHDM inhomogeneities at late times in
the matter era. In the limit
$\eta \gg \eta_{eq}$, the Green functions are dominated by the growing mode,
$\propto a/a_{eq}$.
Hence, the function we would like to solve for is
\begin{equation}
T(k;\eta_i) = \lim_{\eta/\eta_{eq} \to \infty} {a_{eq} \over a}
\widetilde {\cal G}^h(k,\eta,\eta_i).
\end{equation}
This function will be used later to construct the transfer function for the
power spectrum of IHDM density fluctuations induced by cosmic strings.

\subsection{The IHDM sound speed}

The adiabatic sound speed of the self-interacting neutrinos $c_s$, is given by
\begin{equation}
c^2_s = \left( \frac { \partial p } { \partial \rho } \right)_S,
\end{equation}
where $p$ stands for pressure, $\rho$ is the mass density of the neutrinos and the
subscript $S$ on the right hand side means that the entropy per particle $S$,
is constant.

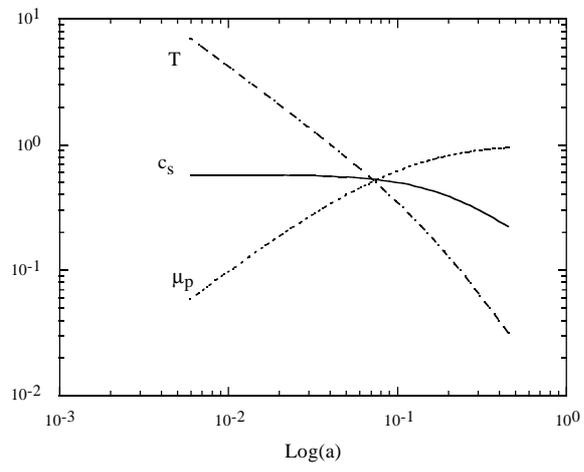
\begin{figure}[t]
\vspace*{7.5cm}
\special{pictfile TmuCs scaled 1000} 
 \caption[]{Plot of $T$, $\mu_p$ (in units of $m$), and $c_s$ as a function of the scale
factor $a$ normalized to unity at $\eta_{\rm eq}$.}     
 \label{fig.1} 
\end{figure}

The neutrinos have Fermi-Dirac distribution function. Therefore, the
neutrino number density $n$, mass density $\rho$ and pressure $p$ are given respectively by: 
\begin{equation}
n = A \, \int_{m}^{\infty} E \, \left( E^2 - m^2
\right)^{1/2}  \frac {1} {\exp\left[(E - \mu_p)/T \right] + 1 } \, dE,
\end{equation}
\begin{equation}
\rho = A \, \int_{m}^{\infty} E^2 \, \left( E^2 - m^2
\right)^{1/2}  \frac {1} {\exp\left[(E - \mu_p)/T \right] + 1 } \, dE,
\end{equation}
\begin{equation}
p = \frac { A } { 3 } \, \int_{m}^{\infty} \left( E^2 - m^2
\right)^{3/2}  \frac {1} {\exp\left[(E - \mu_p)/T \right] + 1 } \, dE,
\end{equation}
where $E = \left( {\bf p}^2 + m^2 \right)^{1/2}$ is the energy, ${\bf
p}$ is the momentum, $m$ is the mass, $T$ is the temperature, $\mu_p \equiv \mu_p(T)$ is
the chemical potential, $A=g /{(2 \pi)}^2$, and $g$ is the spin
degeneracy factor of the neutrinos. Using the constraint 
\begin{equation}
S = \frac { 1 } { T } \left( \frac { p + \rho } { n } - \mu_p \right) = {{7 \pi^4} \over {135
\zeta (3)}},
\end{equation}
where $S$ is the entropy per neutrino, we 
numerically calculated $n$, $\rho$, $p$ and $\mu_p$ as functions of $T$. 
We then obtained the adiabatic sound speed $c^2_s = (\partial p/ \partial
\rho)_S$, and fit the obtained result by:
\begin{equation}
c_s(a) = \frac{ \sqrt{3} } { 3 } \left[ 1 + \left(\frac { a } { \alpha } \right)^2
\right]^{-1/2},
\end{equation}
where $\alpha = 0.19$ is the best fit parameter. In Fig.~1 we represent
the numerical plots of $T$ and $\mu_p$ (in units of $m$), and $c_s$, as
functions of the scale factor $a$, normalized to unity at the epoch of equality between
matter and radiation.

\section{Compensation}
\label{sec.3}

The linear perturbations induced by cosmic strings, are
the sum of initial and subsequent perturbations:
\begin{eqnarray}
\delta_h(k;\eta) &=& \delta_h^I(k;\eta) + \delta_h^S(k;\eta)
\cr\cr &=& 4 \pi (1 + z_{eq})
\int_{\eta_i}^\eta \, d\eta'\, T(k;\eta') 
\widetilde\Theta_{+}(k;\eta').
\end{eqnarray}
The transfer function for the subsequent perturbations, those
generated actively, was obtained in the previous section. To include compensation for the initial
perturbations, $\delta_h^I$, we make the substitution:
\bigbreak
\begin{equation} T(k;\eta) \to \Big( 1 + (k_c/k)^2 \Big)^{-1} \, T(k;\eta),
\end{equation}
where $k_c$ is a long-wavelength cut-off at the compensation scale. This results from the fact
that local physical processes cannot produce perturbations on scales much larger than the horizon
size. Consequently, the power spectrum of density perturbations is bounded by a $k^4$ spectrum on
large scales ($k << k_c$), assuming that background fluctuations are uncorrelated for points which
have never been in causal contact. In the I-model of strings, which most resembles the
Bennett-Bouchet (1990) and Allen-Shellard simulations (1990), $k_c=2 \pi /\eta$.
 
\section{Power Spectrum} \label{sec.gxf}

The analytic expression of Albrecht and Stebbins (1992) for the power spectrum of
density perturbations induced by cosmic strings in a $\Omega =1$ FRW universe with
no cosmological constant is 
\begin{eqnarray}
P(k)
&=&16\pi ^{2}(1+z_{eq})^{2} {(G \mu)}^2  \cr\cr &\times&  
\int_{\eta _{i}}^{\eta
_{o}}d\eta ^{\prime }{\cal F}[k\xi (\eta ^{\prime })/a(\eta ^{\prime })]|T(k,\eta ^{\prime
})|^{2},
\label{pspec}
\end{eqnarray}
where the ``$\rho +3p$'' part of the string stress energy tensor, is modeled by
${\cal F}$ given by 
\begin{equation}  {\cal F}[k\xi /a]=\frac{2}{\pi ^{2}}\beta ^{2}\Sigma\frac{\chi
^{2}}{\xi ^{2}}{\Big(1+2(k\chi /a)^{2}\Big)}^{-1}.
\end{equation}
In these equations, $\eta _{0}$ is the conformal time today, $\eta _{i}$ is the
conformal time when the string network was formed, $\chi$ is the typical curvature
scale of the wakes, 
\begin{equation}
\xi \equiv (\rho _{\infty }/\mu )^{\frac{1}{2}}, \, \, \,  \, \, \, \beta
\equiv {\langle v^{2}\rangle}^{1/2}, \nonumber
\end{equation} 
\begin{equation}
\Sigma \equiv \frac{\mu _{r}}{\mu }\gamma _{b}\beta _{b}+\frac{1}{2\gamma _{b}\beta
_{b}}\Big(\frac{\mu _{r}^{2}-\mu ^{2}}{\mu _{r}\mu }\Big),
\end{equation} 
where $\rho _{\infty }$ is the energy density in long strings, $\mu$ is the
string mass per unit length, $v$ is the microscopic
velocity of the string, $\beta _{b}$ is the macroscopic bulk velocity of the string, $\gamma_{b} =
(1 - \beta_{b}^{2})^{\frac { 1 } { 2 }}$, and $\mu _{r}$ is the renormalized mass-per-unit-length. 
We note that $\Sigma \beta ^{2} \sim 1$ and $\chi /a \sim 2\xi /a\sim
\eta /3$ both in the matter and radiation eras (Bennett and Bouchet 1990; Allen and Shellard
1990) and consequently the structure function ${\cal F}$ is always well approximated by
\begin{equation}  {\cal F}[k\eta ]=\frac{8}{\pi ^{2}} \left[1+\frac { 2 } { 9 } \,
(k\eta )^{2}\right]^{-1}.
\end{equation}
We substitute this in equation (\ref{pspec}) in order to calculate the IHDM power spectrum.

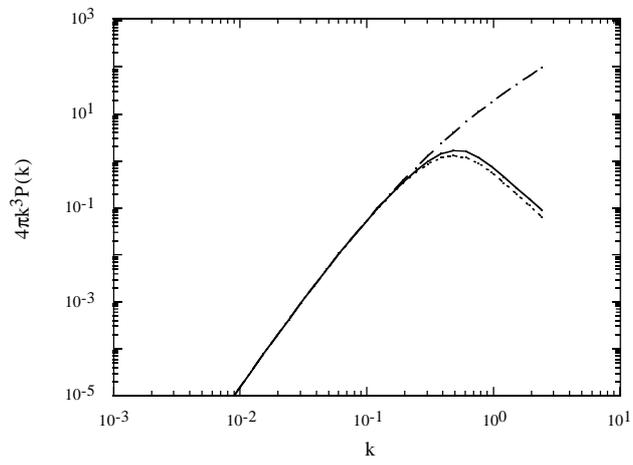
\begin{figure}[t]
\vspace*{7.5cm}
\special{pictfile PS [scaled 1000]} 
 \caption[]{Comparison between the CMB-normalized linear power spectrum of cosmic string seeded
fluactuations with CDM (dot-dashed line), IHDM (solid line) and HDM (doted line) for $\Omega_0=1$
and $h=1$.}     
 \label{fig.2} 
\end{figure}

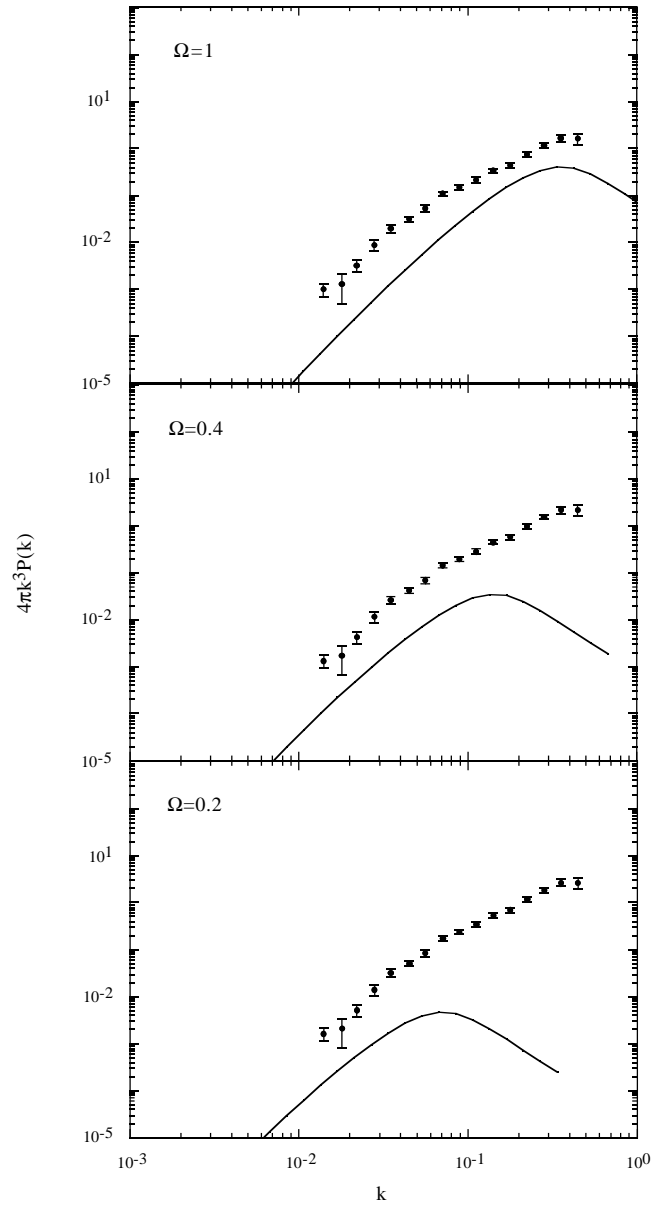
\begin{figure}[t]
\vspace*{17.5cm}
\special{pictfile PowerSpectrum [scaled 1000]} 
 \caption[]{Comparison between the CMB-normalized linear power spectrum of cosmic string seeded
fluactuations with IHDM (solid line) for $h=0.7$ and $\Omega_0=1.0,0.4,0.2$. For $\Omega < 1$
the reconstructed  linear power spectrum has been rescaled by $\Omega^{-0.3}$ (Peacock and
Dodds, 1994).}     
 \label{fig.3} 
\end{figure}

\section{Results}

In Fig.\ 2 we compare the power spectrum of the perturbations generated by cosmic strings for
both CDM (dot-dashed line), HDM (doted line) and IHDM (solid line) for $h=1$ and $\Omega_0=1$.
The curves are normalized to the COBE-DMR observations (Allen el al., 1997). We can see that the
HDM and IHDM power spectra are very similar, the IHDM power spectrum having a slightly  larger
amplitude on small scales. The absence of oscillations reflects the incoherent  nature of string
perturbations which is implicit in the way we add the perturbations generated by the cosmic
strings at different times to calculate the power spectrum. Hence, it seems that the IHDM cosmic
string scenario suffers from similar problems  as the standard HDM scenario for cosmic strings.
We shall discuss these problems next when we investigate other choices for the cosmological
parameters.

\subsection{Open models}
\label{sec.dis}

The generalization of these results for open models of structure formation can be made using 
the same technics employed by Avelino, Martins and Caldwell (1997). In Fig.\ 3 we compare the
power spectrum of the perturbations generated by cosmic strings in the context of IHDM with
the Peacock and Dodds (1994) linear power spectrum reconstrucion inferred from various
galaxy surveys. 

The IHDM power spectrum is again normalized to the COBE-DMR observations
(Allen el al., 1997). We see that in an open universe it is possible to obtain a better 
fit to the shape of the Peacock and Dodds
linear power  spectrum on large scales. However, in this
case perturbations are damped on too large scales which is in conflict with the observational
data. In a flat universe with a cosmological constant the power spectrum requires a slightly lower 
biasing than for an open universe with the same matter density (Avelino, Caldwell and Martins, 1997). However, 
because the shape of the power spectrum density fluctuations induced by cosmic strings remains the same this model 
is also in conflict with observations. 

\section{Conclusions}
\label{sec.dis}

In this paper we calculated the power spectrum of density fluctuations induced by
cosmic strings in the contex of neutrinos with strong self-interactions. We concluded that 
because gravitational instability can only be established on scales smaller than the 
Jeans scale, small scale power is removed relative to the CDM case. In opposition to 
the case studied by Barandela and Davidson (1997), we note the absence of
oscillations, which is due to the incoherent nature of the cosmic strings source. In fact, the 
results obtained for HDM and IHDM are very similar, the IHDM spectrum having a slightly larger
amplitude on small scales. The generalization of these results shows that in an open universe or a flat universe 
with a cosmological constant it is possible to decrease the biasing on large scales by a small 
factor. However, in this case perturbations are damped on too large scales. Hence, we conclude
that it seems difficult to reconcile these results with observations unless there is  some
physical mechanism which is capable of generating a large, scale dependent biasing.

\section*{ACKNOWLEDGMENTS}

P.P.A. is funded by JNICT (Portugal) under the `Program PRAXIS XXI' (grant no. PRAXIS
XXI/BPD/9901/96). We thank Robert Caldwell, Carlos Martins, Jiun-Huei Wu and Paul Shellard
for useful discussions. We thank Centro de Astrof{\' \i}sica da Universidade do Porto (CAUP) for
the facilities  provided.


\begin{thebibliography}{}

\bibitem{AlbrechtStebbins-a} 
 Albrecht A., Stebbins A., 1992, Phys.\ Rev.\ Lett.\ {\bf 68}, 2121
\bibitem{AlbrechtStebbins-b} 
 Albrecht A., Stebbins A., 1992, Phys.\ Rev.\ Lett.\ {\bf 69}, 2615
\bibitem{AllenEtAl} 
 Allen B., Caldwell R.\ R., Shellard E.\ P.\ S., Stebbins A., Veeraraghavan S., 1996,
 Phys.\ Rev.\ Lett.\ {\bf 77}, 3061 
\bibitem{AllenShellard} 
 Allen B., Shellard E.\ P.\ S., 1990, Phys.\ Rev.\ Lett., {\bf 64}, 685
\bibitem{}
 Atrio-Barandela F., Davidson S., 1997, astro-ph/9702236  
\bibitem{Avelino} 
 Avelino P.\ P., 1996, Ph.D. Thesis, University of Cambridge 
\bibitem{AveCal} 
 Avelino P.\ P., Caldwell R.\ R., 1997, in preparation 
\bibitem{AveCalMar} 
 Avelino P.\ P., Caldwell R.\ R., Martins C.\ J.\ A.\ P., 1997, accepted for
 publication in Phys.\ Rev.\ D (see astro-ph/9708057) 
\bibitem{BennettProc} 
 Bennett D.\ P., 1990, High Resolution Simulations of Cosmic String Evolution:
 Numerics and Long String Evolution, in {\it The Formation and Evolution of Cosmic
 Strings}, eds.\ Gary Gibbons, Stephen Hawking, and Tanmay Vachaspati, Cambridge
 Univ.\ Press, Cambridge
\bibitem{BennettBouchet} 
 Bennett D.\ P., Bouchet F.\ R., 1990, Phys.\ Rev.\ D {\bf 41}, 2408 
\bibitem{BouchetProc} 
 Bouchet F.\ R., 1990, High Resolution Simulations of Cosmic String Evolution:
 Small Scale Structure and Loops, in {\it The Formation and Evolution of Cosmic
 Strings}, eds.\ Gary Gibbons, Stephen Hawking, and Tanmay Vachaspati, Cambridge
 Univ.\ Press, Cambridge
\bibitem{}
 Carlson E.\ D., Machacek M., Hall L.\ J., 1992, Ap.\ J., {\bf 398}, 43  
\bibitem{ColombiThesis} 
 Colombi S., 1993, Ph.D. Thesis, UniversitŽ de Paris 7 
\bibitem{}
 de~Laix A.\ A., Scherrer R.\ J., Schaefer R.\ K., 1995, Ap.\ J., {\bf 452}, 495  
\bibitem{}
 Gradwohl B., Frieman J.\ A.,1992,  Ap.\ J., {\bf 398}, 407
\bibitem{Kibble} 
 Kibble T.\ W.\ B., 1976, J.\ Phys.\ A, {\bf 9}, 1387
\bibitem{}
 Kolb E.\ W., Turner M.\ S., 1987, Phys.\ Rev.\ D, {\bf 36}, 2896
\bibitem{}
 Machacek M., 1994, Ap.\ J., {\bf 431}, 41
\bibitem{}
 Nussinov S., Roncadelli M., 1983, Phys.\ Lett.\ B, {\bf 122}, 157
\bibitem{PeacockDodds} 
 Peacock J.\ A., Dodds S.\ J., 1994, MNRAS, {\bf 267}, 1020
\bibitem{}
 Raffelt G.\ C., Silk J., 1987, Phys.\ Lett.\ B, {\bf 192}, 65 
\bibitem{AllenShellardProc} 
 Shellard E.\ P.\ S., Allen B., 1990, On the Evolution of Cosmic Strings, in {\it
 The Formation and Evolution of Cosmic Strings}, eds.\ Gary Gibbons, Stephen Hawking,
 and Tanmay Vachaspati, Cambridge Univ.\ Press, Cambridge 
\bibitem{Traschen} 
 Traschen J., 1985, Phys.\ Rev.\ D, {\bf 31}, 283 
\bibitem{VS90} 
 Veeraraghavan S., Stebbins A., 1990, Ap. J. {\bf 365}, 37 
\bibitem{VilenkinReport} 
 Vilenkin A., 1985, Phys.\ Rep.\ {\bf 121}, 263 
\bibitem{VilenkinShellard} 
 Vilenkin A., Shellard E.\ P.\ S., 1994, Cosmic Strings and other Topological
 Defects.\ Cambrige Univ.\ Press, Cambridge

\end{thebibliography}
\end{document}